\documentclass{jpp}
\usepackage{graphicx}
\usepackage{epstopdf, epsfig}
\usepackage{hyperref}
\usepackage{amsmath,amsfonts,amssymb}
\usepackage{subfigure}
\newcommand*\diff{\mathop{}\!\mathrm{d}}

\shorttitle{Gamma-Diagnostics for Pair Plasma}
\shortauthor{J. von der Linden and others}

\title{Annihilation-Gamma-based Diagnostic Techniques for Magnetically Confined Electron--Positron Pair Plasma}

\author{J. von der Linden\aff{1}\corresp{\email{jens.von.der.linden@ipp.mpg.de}}, S. Ni{\ss}l\aff{1},
  A. Deller\aff{1}, J. Horn-Stanja\aff{1}, J. R. Danielson\aff{2}, M. R. Stoneking\aff{3}, A. Card\aff{4}, T. Sunn Pedersen\aff{5}\footnote{Present affiliation: Type One Energy Group, Madison WI 53703, US},  and E. V. Stenson\aff{1}}

\affiliation{\aff{1}Max Planck Institute for Plasma Physics, Divison E1, 85748 Garching, Germany
\aff{2}University of California San Diego, La Jolla, CA 92093, USA
\aff{3}Lawrence University, Appleton, WI 54911, USA
\aff{4}Technische Universit{\"a}t M{\"u}nchen, 85748 Garching, Germany 
\aff{5}Max Planck Institute for Plasma Physics, Division E4, 17491 Greifswald, Germany}

\begin{document}

\maketitle

\begin{abstract}
Efforts are underway to magnetically confine electron--positron pair plasmas to study their unique behavior, which is characterized by significant changes in plasma time and length scales, supported waves, and unstable modes.
However, use of conventional plasma diagnostics presents challenges with these low-density and annihilating matter-antimatter plasma.
To address this problem, we propose to develop techniques based on the distinct emission provided by annihilation.
This emission exhibits two spatial correlations: the distance attenuation of isotropic sources and the back-to-back propagation of momentum-preserving 2-$\gamma$ annihilation.
We present the results of our analysis of the $\gamma$ emission rate and the spatial profile of the annihilation in a magnetized pair plasma from direct pair collisions, from the formation and decay of positronium, as well as from transport processes.
In order to demonstrate the effectiveness of annihilation-based techniques, we tested them on annular $\gamma$ emission profiles produced by a $\beta^+$ radioisotope on a rotating turntable.
Direct and positronium-mediated annihilation result in overlapping volumetric $\gamma$ sources, and the 2-$\gamma$ emission from these volumetric sources can be tomographically reconstructed from coincident counts in multiple detectors.
Transport processes result in localized annihilation where field lines intersect walls, limiters, or internal magnets.
These localized sources can be identified by the fractional $\gamma$ counts on spatially distributed detectors.
\end{abstract}






\section{\label{sec:intro}Introduction}
There are several efforts underway to magnetically confine cold (0.01 - 10 eV)\citep{} as well as relativistic electron--positron pair plasma \citep{Higaki2010PRE, Stoneking2020JPP, Hicks2019P, vonderLinden2021POP, Peebles2021POP}.
The efforts towards creating magnetically confined cold pair plasma are motivated by the perfect mass symmetry of pairs resulting in a drastic changes in the time- and length-scales as well as to the anticipated mode behaviour \citep{Stenson2017JPP}.
If other symmetry breaking conditions such as species temperature differences can be avoided, the perfect symmetry of pairs will suppress electrostatic instabilities \citep{Helander2014PRL, Mishchenko2018JPP}.
In order to study this behavior in the laboratory, a pair plasma with a unity or small Debye length relative to the plasma size is needed; a $10$ liter plasma size requires $10^{9}-10^{11}$ positrons.
\citet{Pedersen2012NJP} and \citet{Stoneking2020JPP} map out a path towards magnetically confined pair plasma involving accumulating positrons in non-neutral plasma traps from the NEPOMUC positron beam \citep{Hugenschmidt2012NJP}, the world's highest flux positron source, and injecting them (in combination with electrons) into a magnetic confinement geometry suitable for low plasma densities such as a dipole field or a stellarator.
Recently, significant progress has been made towards confining positrons in a permanent magnet dipole trap including lossless injection of a positron beam \citep{Stenson2018PRL}, $>1$ s confinement of positrons \citep{HornStanja2018PRL}, and injection of positrons into the dipole field populated with a dense cloud of electrons (with electron density, $n_{e^{-}} \sim 10^{12}$ $\mathrm{m^{-3}}$) \citep{Singer2021POP}.
\par
Diagnosing a matter-antimatter plasma requires a new set of techniques beyond traditional plasma physics approaches \citep{Hutchinson2002}.
The annihilation of positrons on material surfaces limits the utility of internal probes to situations where termination of the plasma is acceptable, such as setups to verify the injection of positrons into the confinement field \citep{Saitoh2015NJP}.
The lack of coupling between density  and electrostatic potential fluctuations \citep{Stoneking2020JPP} precludes diagnostic techniques of non-neutral plasma.
The low density targeted for positron--electron plasma limits the applicability of electromagnetic-interaction-based diagnostics such as interferometry or Thomson scattering.
With no partially ionized species it will also not be possible to collect passive emission from plasma constituents (although spectroscopy of the neutral bound states of positronium may be possible \citep{Mills2014JPCS}).
Magnetic spectrometers have been used to diagnose the energy distribution of relativistic pair beams \citep{vonderLinden2021RSI}. 
With high magnetic fields and temperatures measuring cyclotron emission may be possible.
However, the most promising diagnostic approaches make use of keV gammas produced by the annihilation of positrons.
This is thanks to the spatial correlations inherent in isotropic and momentum-conserving annihilation emission.
Additionally, while in relativistic pair beams the pair generating target interactions produce bremsstrahlung which obscures the annihilation signal \citep{Chen2012RSI, Burcklen2021RSI}, in low energy positron experiments ($\leq 10$ eV) the high-energy $\gamma$ annihilation signal has a high signal-to-noise ratio. 
\par
The mean expected gamma count rate $C_i$ for one detector $i$ or coincident count rate $C_{ij}$ of two detectors $i$ and $j$, can be modeled as the product of a sensitivity function for the detector(s) $a_i(\vec{x})$ ($a_{ij}(\vec{x})$) and the source distribution $f(\vec{x})$, integrated over the field of view ($\mathrm{FOV}$) of the detector(s),
\begin{equation}
  C_i = \int_{FOV} a_i(\vec{x}) \cdot f(\vec{x}) d\vec{x},
  \label{eqn:detection_equation}
\end{equation}
where the vector $\vec{x}$ defines the 3D coordinate \citep{Defrise2005}.
The sensitivity function $a_i(\vec{x})$ incorporates the detector sensitivity but also scattering effects for the geometry including attenuating materials surrounding the source\footnote{Non-linear effects such as successive counts affecting each other, e.g. due to detector dead-time or random coincidences, have to be modeled separately.}. 
\par The photon counts detected from isotropic radiation sources, such as annihilating positrons, depends on the solid angle $\Omega_i(\vec{x})$ of the source at $\vec{x}$ covered by the detector ($a_i(\vec{x}) \sim \Omega_i(\vec{x})$).
For a given detector (or multiple identical detectors), placed at distances from the source much greater than the spatial extent of each detector, the relative count fraction scales with the inverse of distance (between source at $\vec{x}$ and detector at $\vec{r}_i$) squared ($\Omega(\vec{x}) \propto 1/\vert \vec{x} - \vec{r}_d \vert^2$).
This property is exploited by arrays of uncollimated detectors \citep{Orion1996INIS, Shirakawa2007NIMPR} or equivalently, a single moving detector \citep{Alwars2021SR} to locate individual as well as multiple localized radioactive sources.
\par When annihilation produces two $\gamma$-photons, they have the same $511$-keV energy and propagate nearly $180^\circ$ apart.
Coincident detection of these photons with two detectors indicates that the source likely lies on the line of response (LOR) connecting them.
The field of view is effectively reduced to the line of response ($\mathrm{FOV} \rightarrow \mathrm{LOR}$). 
Detection along multiple intersecting LOR allows triangulation of the positions of the sources.
Gamma detector arrays use coincidences to track several localized sources of annihilation in fluids \citep{Parker1993NIMPR, Parker2002NIMPR, Windows-Yule2022RPP}.  
In magnetized confinement experiments LOR through the magnet and wall could measure radial inward and outward transport resulting in annihilation on material surfaces at known locations.
Coincident count rates on LOR through the confinement volume are effectively line integrations or equivalently Radon transforms of the annihilation source \citep{Radon1917}, $C_{ij} = \int a_{ij}(\vec{x}) f(\vec{x}) dl$, lending themselves to tomographic reconstruction techniques \citep{Maier2018}.
\par
The observation of lossless injection and long-term confinement of positrons in a dipole trap have been based on the interpretation of annihilation detection from two Bismuth germanate (BGO) detectors.
In the injection experiments, positrons annihilated on a target probe after half a toroidal transit \citep{Stenson2018PRL}.
The FOV of a detector was collimated with lead to count gammas originating from the target.
The confinement times were determined by counting either losses or the number of confined positrons as a function of time after injection of a positron pulse \citep{HornStanja2018PRL}.
Losses were measured with an uncollimated detector viewing a large section of the magnet and electrode walls over $10$-ms integration intervals \citep{Saitoh2015NJP}.
At a given time, the positron inventory was measured by counting annihilation after applying a bias potential to localized electrodes, which resulted in the loss of all positrons within one toroidal drift period ($\sim 20$ $\mathrm{\mu}$s).
In all cases the counts had to be averaged over several cycles to achieve acceptable signal-to-noise ratios.
The use of collimated views provides a clear localization of the detected emission but reduces the amount of acquired data.
\par
Upgrades are underway to the dipole confinement experiment \citep{Stoneking2020JPP} that will increase the number of confined positrons and correspondingly the number of annihilations during confinement experiments.
The permanent magnet trap will be replaced with a levitating superconducting coil \citep{Boxer2010NP, Yoshida2010PRL}, providing a $1$ T magnetic field in a cylindrical confinement chamber with a radius of $20$ cm.
A non-neutral buffer-gas trap system \citep{Surko1989PRL} will be installed in the NEPOMUC beam line to accumulate $10^{8}$ positrons and a high-field multi-cell trap is being developed to further increase the accumulation to $> 10^{10}$ positrons \citep{Singer2021RSI}. 
For diagnostics, an array of detectors will be arranged around the confinement volume, increasing the coverage in both solid angle and lines of response.
Pulse-processing hardware will timestamp detections and determine the photon energy absorbed in the detector, allowing for the differentiation between two- and three-$\gamma$ annihilation.
\par
While these annihilation-based techniques are intriguing, annihilation in a matter-antimatter plasma is complex with multiple competing two- and three-body processes contributing to a complicated source function $f (\vec{x})$.
In this paper, we first discuss the various annihilation mechanisms, estimate their rates and spatial extents in order to characterize the source distribution, $f(\vec{x})$.
We then proceed to  characterize the sensitivity function $a(\vec{x})$ of the proposed detector array and demonstrate techniques to diagnose dominant annihilation processes.
\section{\label{sec:ann_rates}Annihilaton processes and rates}
\par
\begin{figure}
  \centering{\includegraphics[width=\textwidth,keepaspectratio]{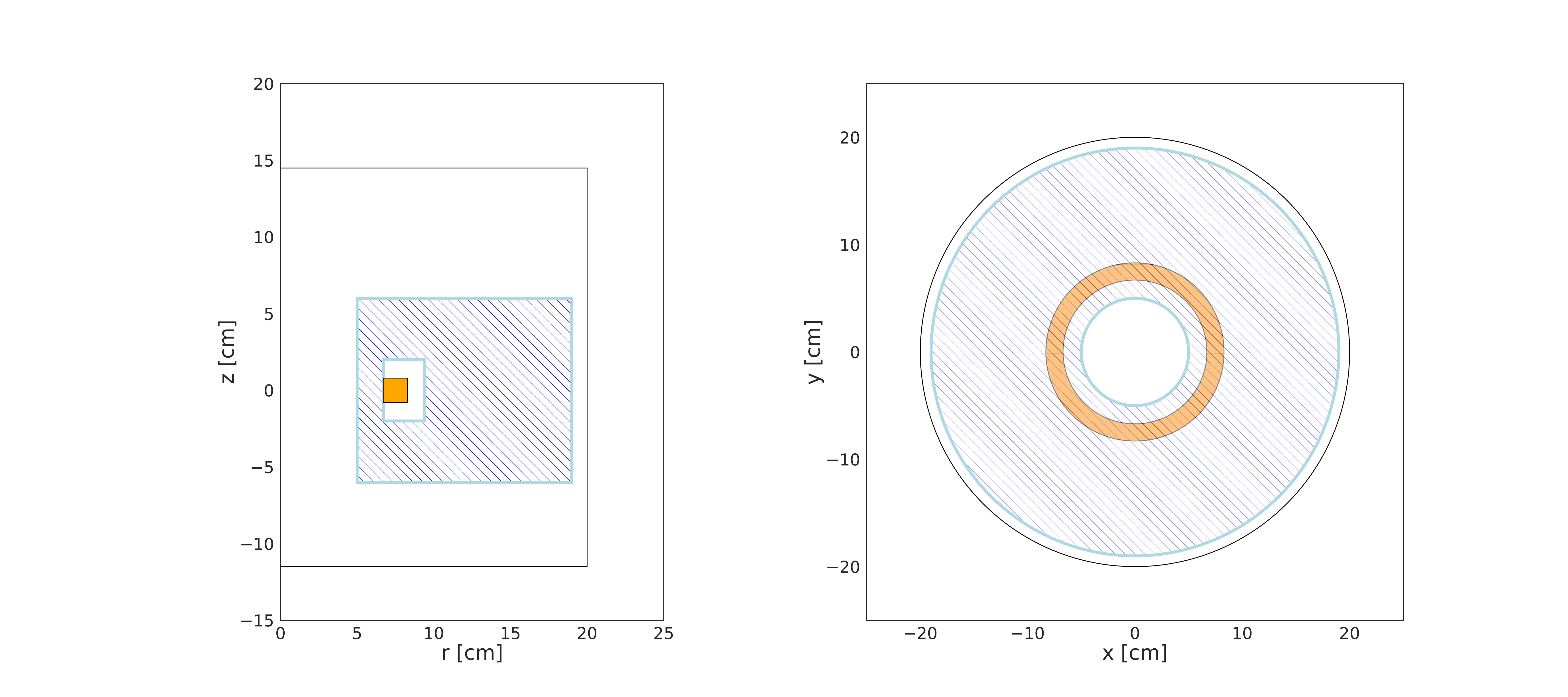}}
  \caption{Simplified geometry of pair plasma in levitating dipole. Floating coil (orange) of $7.5$ cm radius levitates in vacuum chamber (black outline).
The pair plasma is assumed to be confined in a toroid with rectangular cross-section (blue hatched).
The cross-section has a hole where field lines connect to the magnet.
(a) Cross-section. (b) Top view.}
  \label{fig:layout}
\end{figure}
Positrons in a magnetically confined pair plasma annihilate with: free electrons, bound electrons in the background gas, and form short-lived bound states with electrons, positronium (Ps), which eventually annihilate.
Collisions between positrons and other charged or neutral particles transport positrons towards the wall of the confinement chamber or, depending on magnetic geometry, towards exposed magnets, e.g. in the case of a levitating dipole.
\par
To compare the rates and spatial distribution of annihilation we need to choose a parameter range and magnetic confinement geometry.
In this study the density-temperature space considered is in the range $0.01$ eV $\leq T \leq 5$ eV and $10^{11}$ $\mathrm{m}^{-3} \leq \rho  \leq  10^{13}$ $\mathrm{m}^{-3}$.
This discussion uses the levitated dipole experiment as a reference for geometry and plasma parameters (fig. \ref{fig:layout}).
A levitating coil (orange in fig. \ref{fig:layout}) produces a dipole field in a cylindrical chamber (black) with radius $a=20$ cm and height $h=26$ cm.
While the equilbria of magnetized pair plasma have not yet been observed, electron--ion plasmas \citep{Boxer2010NP, Yoshida2013PPCF} as well as non-neutral electron plasmas \citep{Saitoh2010POP} have been confined in levitating dipole fields and there have been theoretical calculations for thermal equilibrium of non-neutral plasma in a dipole field trap \citep{Steinbrunner2023JPP} and maximum entropy states with adiabatic invariant constraints for pair plasma in a dipole field \citep{Sato2023POP}.
This discussion assumes a simplified rectangular cross-section for the floating coil and plasma.
This plasma profile is unphysical, but here we are concerned with identifying gross annihilation profiles. 
The cross-section of the coil is square with $1.6$ cm sides centered at a radius of $7.5$ cm.
The pair plasma is assumed to be confined within a rectangular cross-section (hatched blue) extending radially $5<r_p<19$ cm and axially $-6<z_p<6$ cm.
The pairs on field lines intersecting the magnet are assumed to be lost, resulting in a plasma-free shadow around the magnet extending radially from $6.7 <r_s< 9.4$ cm and axially $-2 < z_s < 2$ cm.
\par
\citet{Stoneking2020JPP} discussed positron annihilation with free and bound electrons as well as due to Ps formation in a magnetized pair plasma in terms of their effect on the lifetime of the pair plasma.
Under ultra-high-vacuum conditions, when direct annihilation with bound electrons on neutrals and charge-exchange become negligible the main contributions to annihilation were found to come from Ps formation via radiative recombination and subsequent annihilation and direct annihilation with free electrons. 
At temperatures of several eV and higher, Ps formation through charge-exchange with residual gas atoms may dominate the other processes but we will ignore this case here.
\par
The mechanisms discussed so far originate in the bulk of the pair plasma.
In a multi-species plasma there is transport towards boundaries such as the walls and magnet.
Most positrons annihilate once they reach solid boundaries.
Diffraction from low-energy positrons hitting solid surfaces is limited to no more than $\sim 10$\% of the incoming positrons \citep{Rosenberg1980PRL, Schultz1988RMP} 
Here, we consider transport driven by Coulomb collisions and scattering off neutrals.
In pair plasmas there is also the possibility of Ps mediated transport where Ps forms, drifts across the magnetic field and ionizes either through collisions or fields, as has been studied for the case of antihydrogen in positron--antiproton traps \citep{Jonsell2009JPB, Jonsell2016JPB}. 
Predicting transport processes in plasmas is difficult, but models can give estimates that can be checked by experiments.
Scattering off neutrals is thought to be the main loss process in the low-density positron confinement experiments \citep{HornStanja2018PRL}. 
The measurements from these experiments can be scaled to the levitating dipole geometry.
In a strongly magnetized plasma, where the Debye length $\lambda_D$ is longer than the Larmor radius $r_L$, collisions differ from classical plasma collisional theory.
Due to the low densities, pair plasma will be strongly magnetized \citep{Stenson2017JPP}.
Theory \citep{Dubin1998POP, Dubin1997PRL} and observations \citep{Anderegg1997PRL} in non-neutral plasma suggest the diffusion coefficient is enhanced for collisions with an impact parameter larger than the Larmor radius, $\rho > r_L$. 
For both transport processes, the diffusion rate is taken to be the annihilation rate.
\par
\begin{figure}
  \centering{\includegraphics{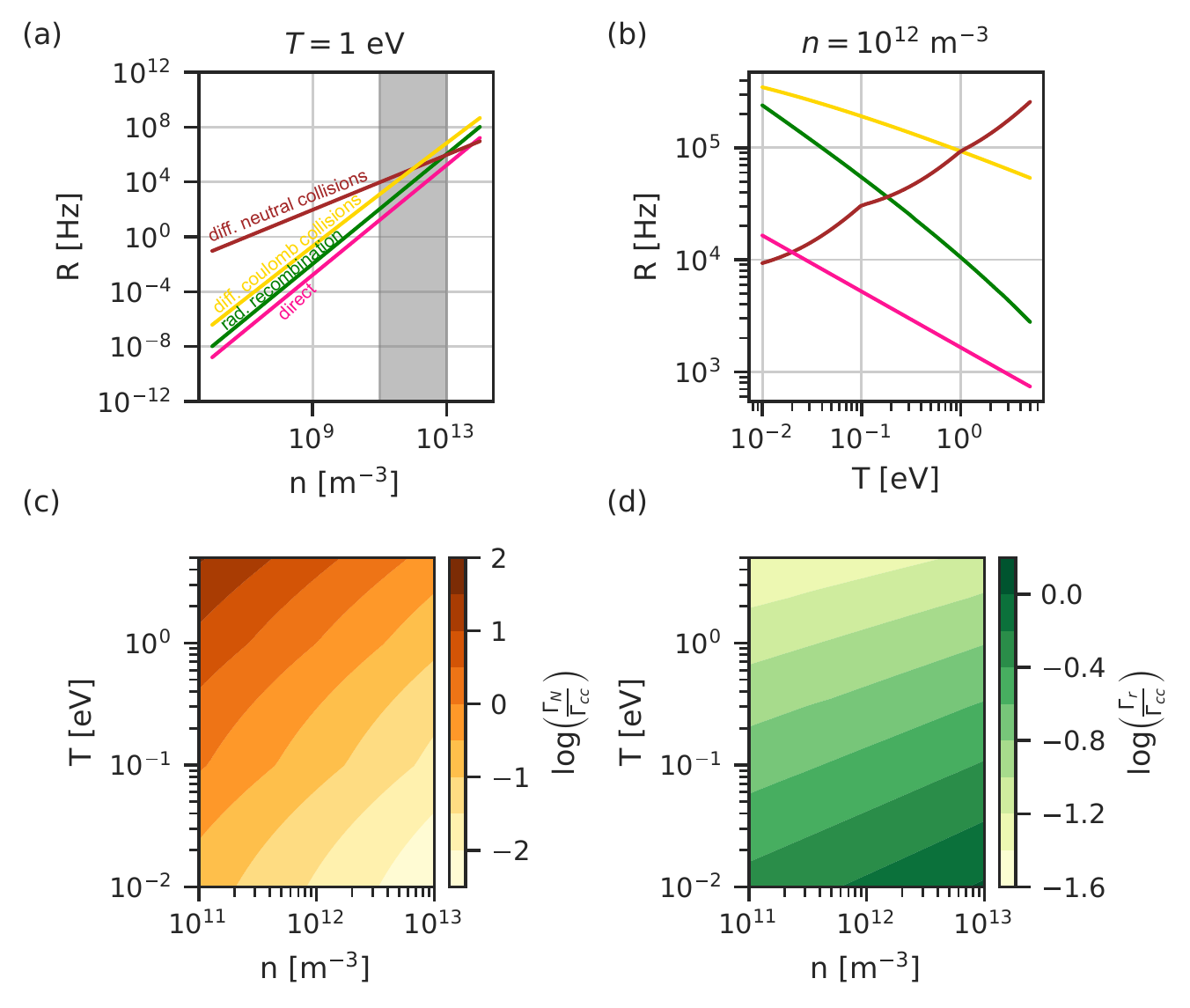}}
  \caption{Annihilation rates $R$ due to direct pair collisions (pink), radiative recombination (green), Coulomb collision diffusion (yellow) and neutral collision diffusion (brown) in a 12-liter pair plasma in the simplified dipole confinement geometry. 
    a) Density dependence of annihilation rates $R$ of a pair plasma with temperature $1$ eV.
The gray region marks the targeted densities for low energy pair plasma experiments.
    b) Temperature dependence of annihilation rates of a pair plasma with density $10^{12}$ $\mathrm{m^{-3}}$.
    c) Ratio of neutral collision diffusion to Coulomb collision diffusion over density-temperature space.
    d) Ratio of the rate of radiative recombination to the rate of Coulomb collision diffusion over density-temperature space.} 
  \label{fig:rates}
\end{figure}
\par
Fig. \ref{fig:rates} (a,b) shows the annihilation rates due to radiative recombination (green), direct (pink), Coulomb collision (yellow) and neutral collision (brown) processes in a magnetized pair plasma as a function of density and temperature (see the Appendix for rate equations).
The annihilation rate plotted is for all positrons in the volume $R=\Gamma N_{e^+}$, $\Gamma$ is the annihilation rate of a single positron and $N_{e^+}$ is the initial number of confined positrons ($R$ is equivalent to the volume integral of source function over all space $\int f(\vec{x}) dV$).
$R$ represents an instantaneous rate of annihilation and gives a sense for how large the emission signal is from the plasma; this rate declines as the positron number depletes.
However, since annihilation has been found to constrain the pair plasma lifetime to $>10^3$ s \citep{Stoneking2020JPP}, $R$ approximates the rate during the first seconds or minutes of confinement and we will not consider the time dependence of the source distribution. 
Below the targeted pair plasma regime densities ($n<10^{11}$ $\mathrm{m^{-3}}$), these plasmas are transport limited; diffusion to material surfaces due to neutral collisions dominates the other annihilation processes by several orders of magnitude.
Transport due to Coulomb collisions as well as the rates of radiative recombination and direct annihilation increase with density, faster than transport due to neutral collisions.
While the ratio between direct annihilation and radiative recombination is independent of density, the positron density does affect their respective ratios to transport processes.
Diffusion due to Coulomb collisions will overtake diffusion due to neutral collisions around $n \sim 10^{12}$ $\mathrm{m^{-3}}$ and radiative recombination around $n \sim 9 \cdot 10^{12} \mathrm{m^{-3}}$.
This suggests that positron annihilation lifetime spectroscopy measurements \citep{Cassidy2006APL} of a $n=10^{13} \mathrm{m^{-3}}$ pair plasma may see two distinct loss regimes as the plasma decays.
Annihilation of free positrons with electrons results in the production of two gammas most of the time.
\par
\begin{figure}
  \centering{\includegraphics[width=\textwidth,keepaspectratio]{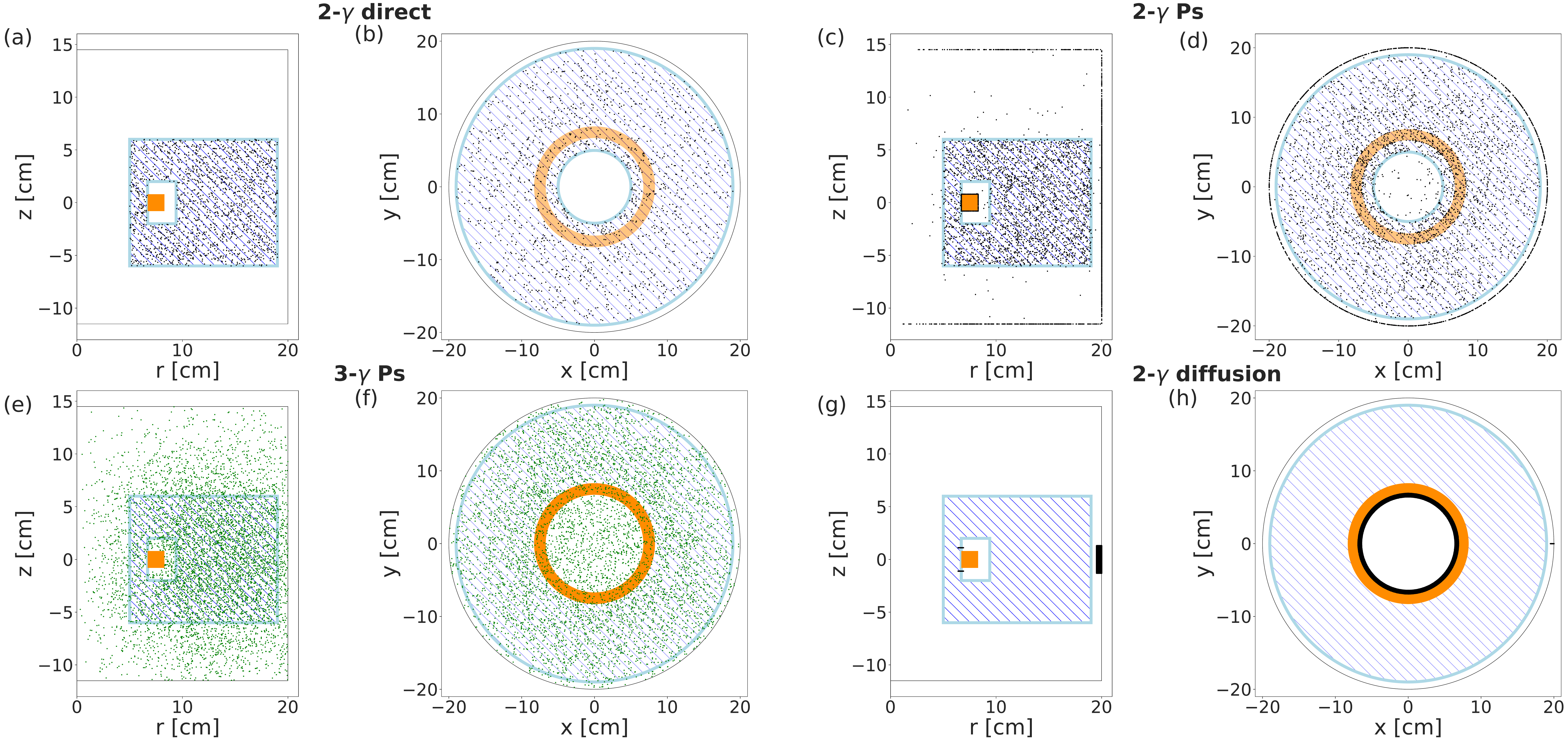}}
  \caption{Spatial distribution of annihilation events in one second in $1$ eV, $12$ liter-pair plasma with density $10^{12}$ $\mathrm{m^{-3}}$ magnetically confined in dipole field of levitating dipole as shown in fig. \ref{fig:layout}:
    (a, b) direct annihilation events between positrons and free electrons resulting in 2-$\gamma$ emission,
    (c, d) 2-$\gamma$ decays of positronium,
    (e, f) 3-$\gamma$ decays of positronium,
    (g, h) 2-$\gamma$ emission from annihilation of positrons diffusing from the plasma to magnet and limiter.}
  \label{fig:spatial}
\end{figure}
At the assumed temperatures and densities, the most significant Ps formation channel is radiative recombination.
The lifetime and decay of Ps depends on the spin of the bound particles \citep{Ore1949PR, Deutsch1951PR}.
Parapositronium (pPs) has antiparallel spins and its ground state decays into two gammas with an mean lifetime of $125$ ps.
Orthopositronium (oPs) has parallel spins and its ground state decays into three gammas with a mean lifetime of $142$ ns \citep{Vallery2003PRL}.
With a $1$ eV temperature the ground state of pPs (oPs) can travel $5$ $\mathrm{\mu m}$ ($6$ cm) in its mean lifetime at the most probable speed ($\sqrt{kT/m_e}$).
There is also a small probability of creating $2n$ or $2n+1$ photons, although the branching ratio for the 4 and 5 gamma decays is on the order of $10^{-6}$ and declines further for higher $n$ \citep{Karshenboim2003IJMP}.
For unpolarized positrons $1/4$ of the Ps formation will be pPs and $3/4$ oPs.
Ps may form in excited states with probabilities and lifetimes discussed in the Appendix and Appendix tables \ref{tab:pPs_appendix} and \ref{tab:oPs_appendix} \citep{Gould1989TAJ, Gould1972AP, Alonso2016PRA, Cassidy2018EPJD}.
Magnetic fields can lead to Zemann mixing of singlet and triplet states, which will reduce oPs lifetimes, however, we do consider this effect \citep{Deutsch1951PR2, Alonso2015PRL}.
Ps propagates freely at the chosen velocity until, either the end of its lifetime or it intersects a solid object, i.e. the wall or magnet.
The annihilation signal and location is determined by which state of Ps forms and where the Ps 'walks to'.
To model the spatial distribution of annihilation due to Ps formation, we
\begin{enumerate}
\item randomly distribute formation events over the uniform plasma volume.
\item determine the energy state and the corresponding lifetime (or for higher energy state the total lifetime of the state-cascade) using the lifetimes in tables \ref{tab:pPs_appendix} and \ref{tab:oPs_appendix}.
\item pick each of the three velocity components from a normal distribution centered at $0$ with $\sigma=\sqrt{kT/(2m_e)}$.
\item propagate the Ps along its velocity direction using a $1$ mm step size according to its lifetime and speed.
\item check for intersections with solid objects.
\end{enumerate}
The resulting 3-$\gamma$ signal from oPs is volumetric extending throughout the chamber (fig. \ref{fig:spatial} e, f).
There is a gradient in 3-$\gamma$ source density outside the plasma volume.
oPs intersecting the wall or magnet can interact with solids in multiple ways including pick-off and quenching to para-positronium that lead to fast decay and enhanced 2-$\gamma$ decay probabilities \citep{Cassidy2018EPJD, Schoepf1992PRA, Coleman2002ASS}.
We assume all wall and magnet intersections to contribute to the 2-$\gamma$ signal along with pPs decays (fig. \ref{fig:spatial} c, d).
The 2-$\gamma$ signal from Ps is confined essentially to the plasma volume with the exception of longer lived excited pPs states that can drift out and the oPs that reach the wall and magnet.
The transport results in a localized annihilation signal from the magnet and from a narrow, $\sim 2$ cm in axial extent, azimuthal ring where the field lines intersect the wall.
We assume the transport has no inward/outward preference so that half of the annihilation occurs on the magnet and half on the wall.
The signal can be made more localized if a circular limiter ($1.3$ cm radius, $5$ mm in front of wall at $y=0$ and positive $x$) is introduced (fig \ref{fig:spatial} g, h).
\begin{figure}
  \centering{\includegraphics[width=\textwidth,keepaspectratio]{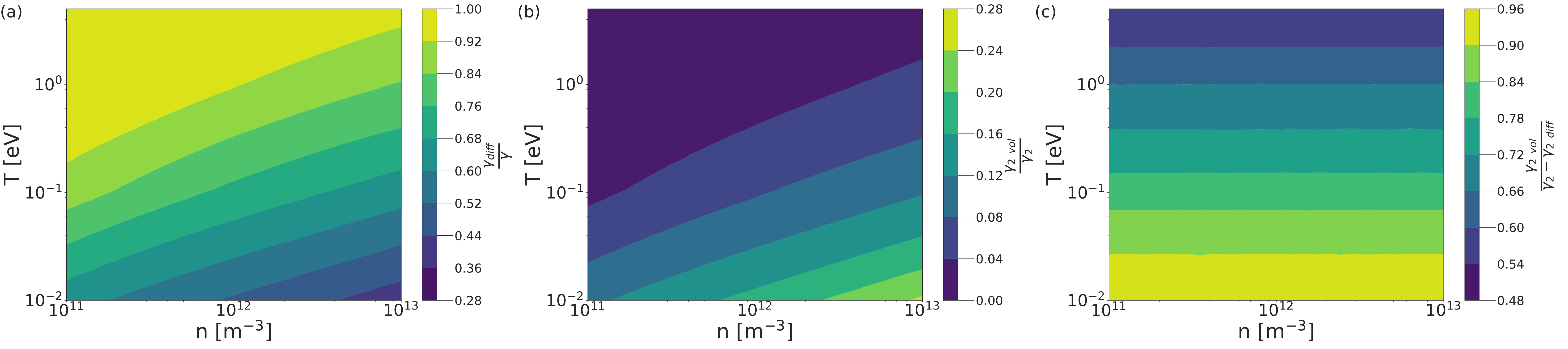}}
  \caption{Emission fractions. 
    (a) Ratio of total number of photons emitted due to diffusion ($\gamma_{diff}$) to all photons emitted ($\gamma$). These ratios account for 2-$\gamma$ and 3-$\gamma$ emission.
    (b) Ratio of volumetric 2-$\gamma$ ($\gamma_{2vol}$) photons to total 2-$\gamma$ photons emitted ($\gamma_2$).
    (c) Ratio of volumetric 2-$\gamma$ ($\gamma_{2vol}$) photons to total 2-$\gamma$ photons emitted minus the diffusion photons ($\gamma_2 - \gamma_{2diff}$).
  }
  \label{fig:simulated_emission_fractions}
\end{figure}
\par
Examining the photon counts further constrain the source distribution $f(\vec{x})$.
Photon counts can be differentiated between the count of all photons, $\gamma$, and the count of photon pairs from 2-$\gamma$ emission, $\gamma_2$.
The latter can be diagnostically identified by their distinct energy signature (511 keV).
Another classification is in terms of the photon origin, denoted by subscripts: $\gamma_{vol}$ for photons emitted from volumetric sources, $\gamma_{bds}$ for photons emitted from Ps hitting boundaries, and $\gamma_{diff}$ for photons emitted when diffusing positrons hit the narrow field-line intersections of the magnet and wall or limiter.
The latter two origins only contribute to the $\gamma_2$ count since in our model annihilation on solids results in 2-$\gamma$ events.
The majority of the photons emitted originate from diffusion for much of the parameter space, making the quantification of diffusion processes a promising diagnostic aim (fig. \ref{fig:simulated_emission_fractions} a).
In dense and cold pair plasma $\gamma_{2vol}$ can exceed 20\% the total $\gamma_2$ but for a large portion of the density-temperature space, the fraction is less than 1\% (fig. \ref{fig:simulated_emission_fractions} b).
2-$\gamma$ emission can be detected by coincidence which is highly localized to the magnet and limiter.
A suitable arrangement of detectors can create LOR that do not cross the diffusion emission regions.
These LOR will only cross a small fraction of the wall and a large fraction of the volume.
The ratio of volumetric 2-$\gamma$ photons to 2-$\gamma$ photons emitted at the boundaries excluding the diffusion photons, $\gamma_{2vol}/(\gamma_2 - \gamma_{2diff})$, stays above 40\% throughout the density-temperature space (fig. \ref{fig:simulated_emission_fractions} c).
The volumetric and localized signals indicate that even with multiple overlapping processes we can likely untangle their contributions.
There are three signals that are of particular interest:
\begin{enumerate}
\item Transport provides emission that is strongly localized to the magnet and wall section or limiter and that has the dominant photon count for much of the parameter space.
The magnitude of this signal is directly related to the physics of transport/diffusion processes.
The strong localization lends itself to a diagnostic method exploiting distance-attenuation.
\item The volumetric 2-$\gamma$ emission that can be filtered due to its distinct energy.
This emission is due to direct annihilation and Ps formation (pPs) which are both related to the density and temperature profiles of the pair plasma. The volumetric 2-$\gamma$ signal is dominated by the transport emission which is 2-$\gamma$ as well.
A suitable choice of detector positions could have LORs with good sampling of both, allowing for tomographic reconstruction.
\item The 2-$\gamma$ signal is localized and related to the positronium (oPs) formation and thermal drift.
Diagnosing this signal with LORs with a long path along the wall may help disentangle the contribution of Ps formation and direct annihilation in signal 2.
\end{enumerate}
\par
\section{\label{sec:array} Gamma-detector array sensitivity}
Here we introduce the gamma detector array, evaluate its sensitivity function $a(\vec{x})$ of equation (\ref{eqn:detection_equation}), quantify its time and energy resolution, and discuss the effect of these quantities on measurement capabilities.
Radioisotope sources are used as effective point sources of emissions to characterize detection systems.
For single photon counting from a point-source of emission ($f(\vec{x}) \rightarrow R\delta(x)$) we approximate the integral over the FOV as the multiplication of the solid angle of the source covered by the detector(s) $\Omega$ with the efficiency factor $\eta$, which includes the detector efficiency as well as all other physics such as attenuation and scattering over all space, 
\begin{equation}
  C_i = \int_{FOV} a_i(\vec{x}) R \delta(\vec{x}) d\vec{x} \sim \Omega_i(\vec{x}) \eta_i(\vec{x}) R.
  \label{eqn:sensitivity_equation}
\end{equation}
In practice, $\eta$ is determined for both the total counts of a detector and the counts in the photo-peak (with subscript $pp$) as the provenance of the latter as non-scattered emission is more certain.
We now proceed to evaluate the solid angle coverage $\Omega_i$ based on the detector geometry and use reference $^{22}\mathrm{Na}$, $\beta+$ emitters, to determine $\eta$. 
\begin{figure}
  \centering{\includegraphics[width=\textwidth,keepaspectratio]{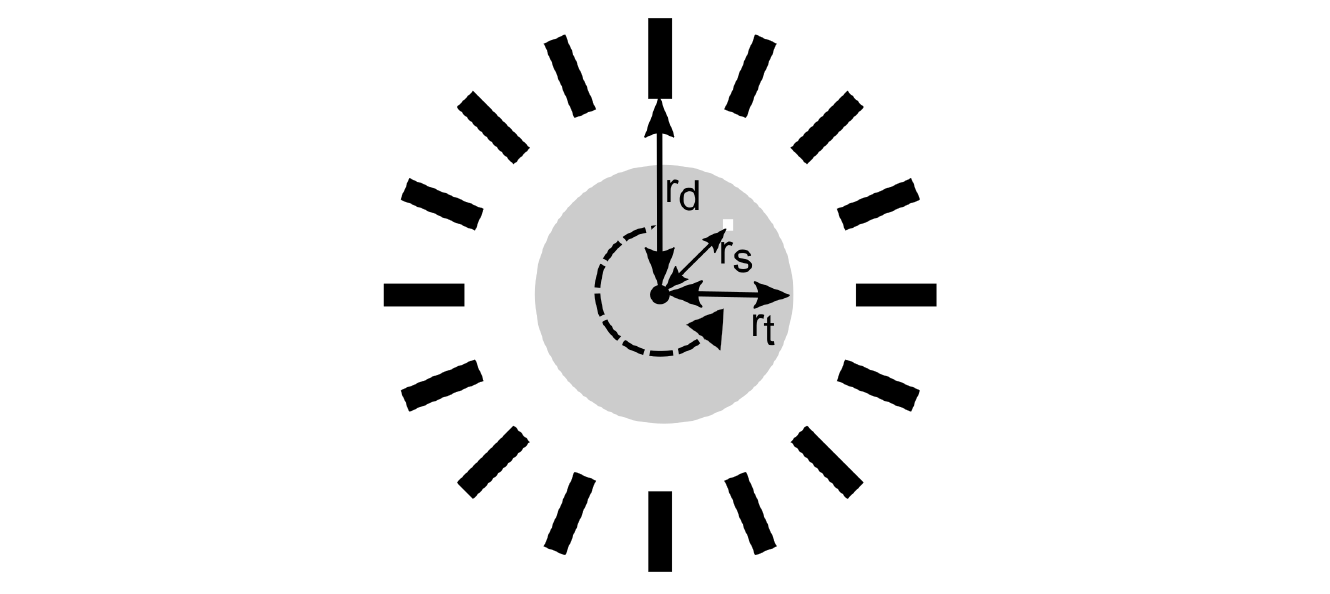}}
  \caption{Test setup imitating annihilation in a toroidal magnetic confinement geometry.
    Sixteen BGO detectors are equally spaced (every $22.5^\circ$) at $33$ cm radius ($r_d$) around a $^{22}\mathrm{Na}$ source (white square at $r_s$) placed on turntable with a 22.5cm radius ($r_t$).}
  \label{fig:turntable}
\end{figure}
\begin{figure}
  \centering{\includegraphics[width=\textwidth,keepaspectratio]{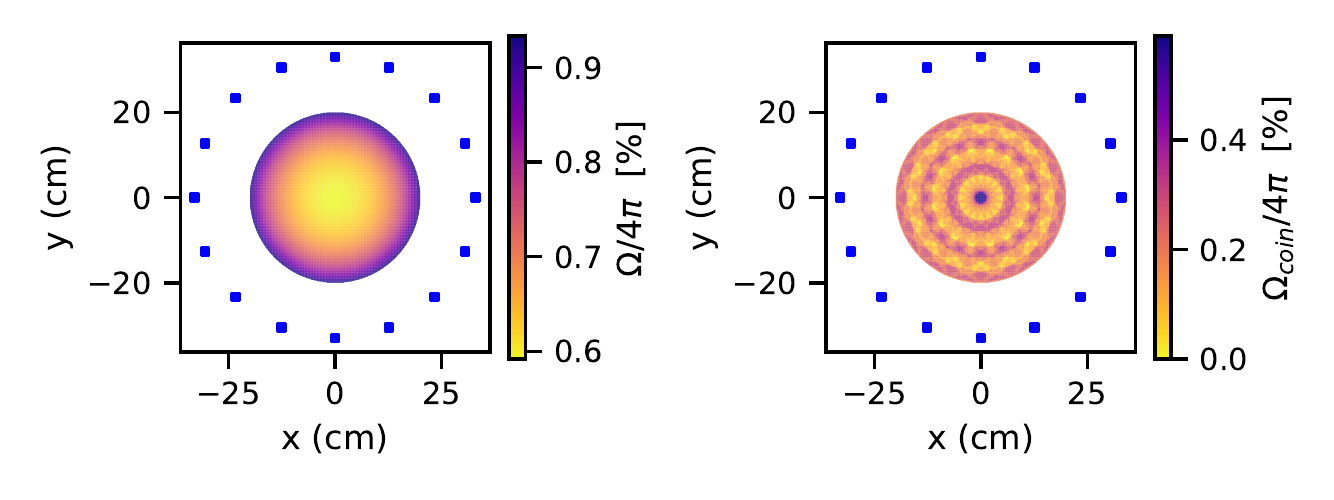}}
  \caption{$4 \pi$ coverage of the 16 detectors (blue squares) arranged in a $33$ cm radius circle.
    a) Sum of solid angle coverage of all detectors for single photons emitted by point source located inside a radius $20$ cm.
    b) Sum of solid angle coverage for two photon coincidence emitted by a point source located inside a radius $20$ cm.}
  \label{fig:solid_angle}
\end{figure}
\par
We use a test setup (fig. \ref{fig:turntable}) with 16 BGO (Scionix 25B25/1M-HV-E2-BGO-X2) detectors arranged in a circle with radius $r_d=33$ cm, which could also fit the 48 detectors ($16.7$ x $3.97$ cm) envisaged for the pair plasma experiments.
Each detector consists of a cylindrical BGO crystal with $2.54$ cm diameter and $2.54$ cm length.
The solid angle coverage of a detector to a point-source is given by \citep{Knoll2010}
\begin{equation}
  \Omega = 2\pi \left( 1 - \frac{\ell}{\sqrt{\ell^2 + \alpha^2}} \right),
  \label{eqn:solid_angle}
\end{equation}
where $\ell=\lvert \vec{r}_i - \vec{x} \rvert$ is the distance between the detector and the source and $\alpha$ is the radius of the scintillator.
Fig. \ref{fig:solid_angle} a) shows the solid angle coverage of the detector arrangement, summing the solid angle coverage of all 16 detectors for point source positions on a $1$-mm grid.
The $4\pi$ coverage varies from $0.6\%$ to roughly $1\%$ at the edges.
Fig. \ref{fig:solid_angle} b) shows the solid angle coverage of the pairs of detectors forming lines of response.
For coincident detection the solid angle for each source point is determined by the detector furthest from the source.
The maximum solid angle coverage for coincidences is the center where the most (8) lines of response meet.
There are several locations with no lines of response and consequently no solid angle coverage.
\par
The detection efficiency as well as non-linear aspects of the response i.e. the rate of false coincidences and missed counts, are influenced by the hardware.
Scintillation in the detectors is converted to electrical pulses with photo-multipliers (Hamamatsu 1924A) and preamplifiers with heights proportional to the absorbed photon energy.
The output pulse from the preamplifiers has a rise time of $140$ ns and a decay time of $1$ $\mathrm{\mu s}$.
FPGA based multi-channel analyzers (CAEN V1730S) digitizes all detector outputs to 14 bit resolution at 500MS/s.
The FPGA timestamps the $50\%$ of peak amplitude point  of each pulse with a digital implementation of a constant-fraction (CFD) trigger and determines the pulse height by digitally integrating a set gate of $150$ ns before and $1850$ ns after the trigger.
During the decay of the pre-amplifier the signal remains above the threshold of the CFD trigger, resulting in a dead time $t_D \sim 4$ $\mu s$.
The fraction of the measured rate to the true rate $R_m/R_t$ can be estimated \citep{Knoll2010} as $R_m/R_t = 1 - R_m t_D$.
Missed events due to dead-time will be significant and need to be accounted for, as the missed counts start to exceed 1\% of the measured rate when $R_m > 2.5 \cdot 10^3$ Hz.
This dead-time does not affect coincidence measurements as these occur on two separate detectors.
\begin{figure}
  \centering{\includegraphics{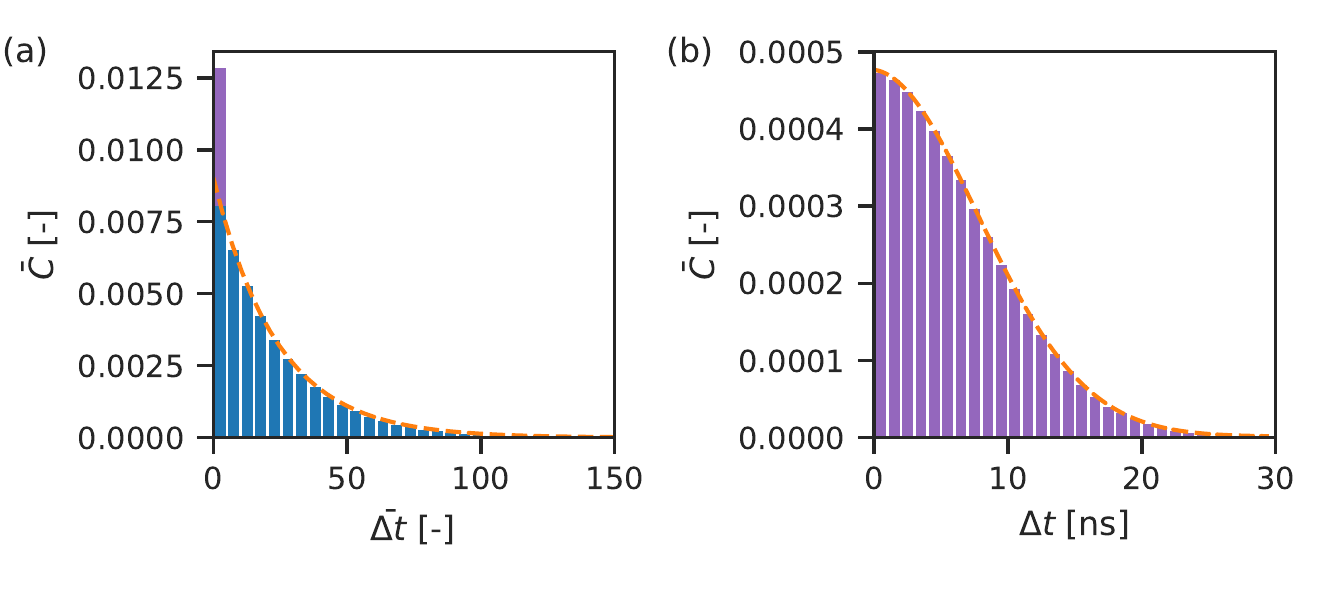}}
  \caption{Frequency of intervals between successive detections on a) total rate and b) coincidence timescale. The count rates and intervals (in (a) but not (b) are normalized by the source activity $\bar{C}=C/R$ and $\bar{\Delta t}=\Delta t \cdot R$, where $R=35$ kBq.
      On long timescales the interval distribution fits an Erlang distribution.
      On nanosecond timescales the distribution of intervals fits a Gaussian distribution with a standard deviation of $\sigma=8$ $\mathrm{ns}$.}
\label{fig:time_intervals}
\end{figure}
\par
We measure the time resolution in order to estimate the rate of false coincidences.
Fig. \ref{fig:time_intervals} a) shows the time intervals between consecutive detection events by the 16 detectors when a $^{22}\mathrm{Na}$ source is placed in the center.
The count rate and time are normalized by the activity of the source. 
The distribution of intervals between events for all detectors fits an Erlang distribution except for the leftmost bin which is over-populated due to coincident detections between pairs of detectors for 2-$\gamma$ annihilations.
Binning for these shortest time intervals reveals that the coincident intervals fit a Gaussian distribution with standard deviation of $8$ ns which is the time response of the detection system.
We treat detections within three standard deviations as coincident, giving a coincidence window $\tau=24$ ns.
The FPGA has been shown to be able to timestamp the square pulses from a delay generator (SRS DG645) to the accuracy of the generator ($1$ ns) so the time response is dominated by the electronics of the BGO detector package.
The fraction of false coincidences can be estimated \citep{Parker2002NIMPR} as $R_{fc}/R_m \sim 2 \tau R_m$.
 $R_{fc}/R_m \sim 1\%$ with $R_m=2\cdot 10^5$ Hz; given the solid angle coverage (fig. \ref{fig:rates} a), the predicted rates of annihilation should result in few false coincidences.
\begin{figure}
  \centering{\includegraphics{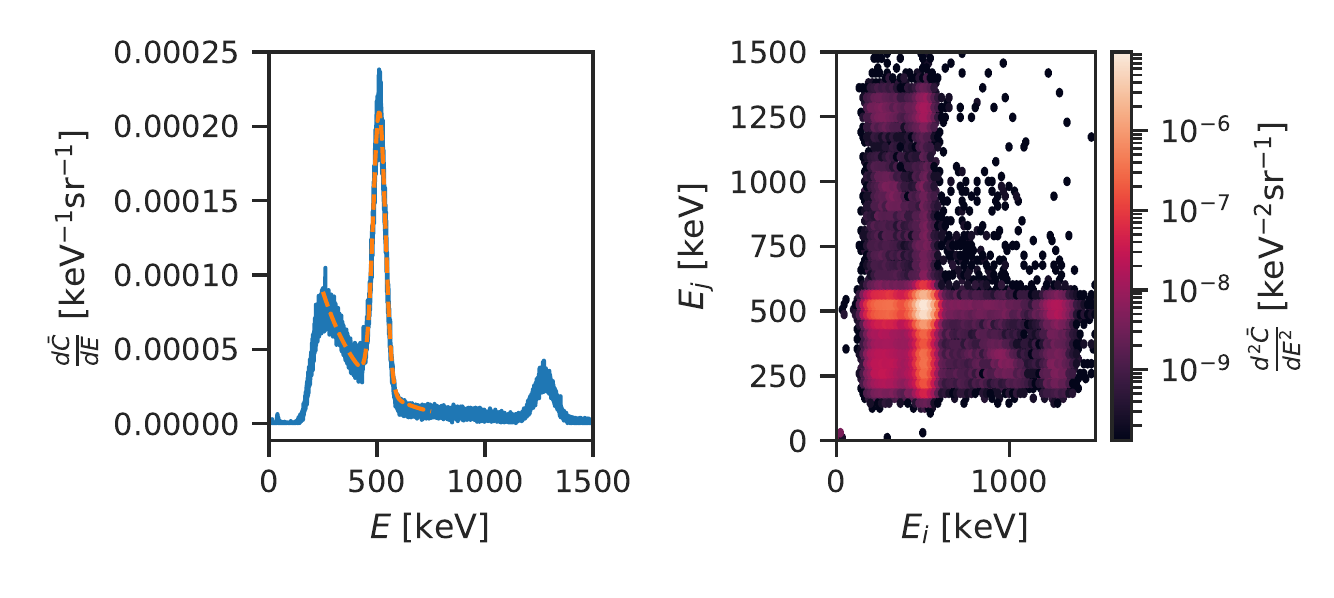}}
  \caption{Energy spectrum from 16 BGO detectors forming a 33cm radius circle around a $\mathrm{^{22}Na}$ source.
    The count rate is normalized by the source activity $\bar{C}=C/R$, where $R=35$ kBq.
  a) Energy spectrum of single photon detections. 
  The spectrum around the $511$-keV photo-peak can be fit by a Gaussian and an exponential (dashed orange).
  b) Energy spectrum of coincident detections on two detectors $i,j$ within $\tau=24$ ns.
  }
  \label{fig:energy_spectra}
\end{figure}
\par
Characterizing the energy resolution of the detector array allows us to estimate how well we can filter for $511$-keV photons and how well we can relate counts to the annihilation rate.  
Fig. \ref{fig:energy_spectra} a) shows the energy spectrum of single detections calibrated with the $^{22}\mathrm{Na}$ peaks at $511$ keV and $1274.5$ keV.
The $511$-keV photo-peak can be fitted by a Gaussian distribution on top of a continuum fitted with an exponential decay (dashed orange) \citep{Knoll2010}.
The FWHM of the $511$-keV annihilation peak is 13\% for gamma spectra acquired in this study, corresponding to a 66keV energy resolution \citep{Karwowski1986NIMPR}.
Fig. \ref{fig:energy_spectra} b) shows the energy spectrum of coincident detections within 24 ns on two detectors $i$ and $j$.
$^{22}Na$ emits 1274.5 keV photon within picoseconds of the positron emission so there can be coincidences between the $511$-keV photons from 2-$\gamma$ annihilation, as well as the $1274.5$ keV photons and the partial absorption of photons due to Compton scattering.
\par
We measure $\eta(\vec{x})$ by comparing the experimental counts from three $^{22}\mathrm{Na}$ sources with equation ($\ref{eqn:sensitivity_equation}$) and taking into account that $f$ is the known source activity adjusted for the photons emitted per decay which is $0.999+1.798$ for all counts and $1.798$ for 511-keV photon peak counts \citep{Delacroix2002}.
$\eta({\vec{x}})$ depends logarithmically on the distance between the source and the detector.
For positions on the turntable, $\eta$ varies between 6 and 7 and $\eta_{pp}$ varies between 0.4 and 0.45. 
In a pair plasma experiment the stainless steel chamber walls and other components will attenuate radiation, necessitating care in calibrating the detection system for a spatially varying factor $\eta(\vec{x})$.
\section{Diagnostic Methods \label{sec:methods}}
\subsection{Distance-attenuated photon counting of transport \label{sec:single_photon_couting}}
\begin{figure}
  \centering{\includegraphics{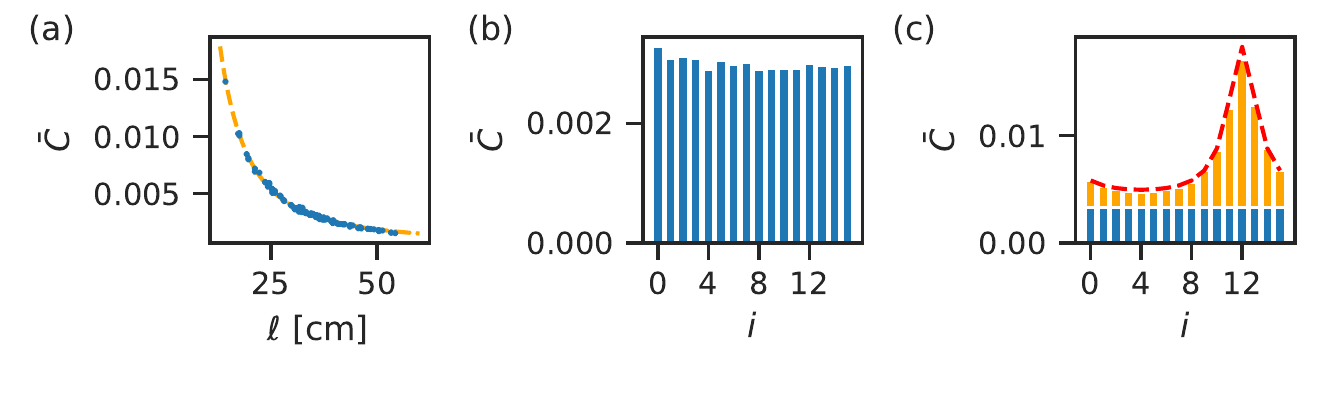}}
  \caption{Identification of localized $\gamma$ source off-axis of an axisymmetric distribution of $\gamma$ emission as an approach for identifying pair plasma diffusion onto a limiter.
(a) Calibration of distance-attenuated photon count rate.
Blue dots are counts per second recorded on detectors a distance $\ell$ from the source.
The measurements fit equation (\ref{eqn:fit_solid_angle}) (dashed orange).
(b) The counts rate on each of the 16 detectors recorded with a $\gamma$ emission distribution $f(x)=\delta(r-r_0)$, with $r_0=7$ cm. 
(c) The counts rate on 16 detectors with a $\gamma$ emission distribution $f(x)=\delta(x=0)\delta(y-y_o) + \delta(r-r_0)$, with $y_0=-20$ cm.
The expected counts for a point-source at $y=-20cm$ is shown in dashed red.}
\label{fig:distance_attenuated_counts}
\end{figure}
Fig. \ref{fig:spatial} (g,h) shows that diffusion in a pair plasma could result in a ring of annihilation on the magnet and localized annihilation on a limiter.
A distance-attenuation calibration of the gamma-detector array can identify the diffusion emission on the limiter.
By placing a $^{22}\mathrm{Na}$ source at 8 different radii, a count function can be fitted to the measurements at each detector    
\begin{equation}
  \bar{C}_i = A \left( 1 - \frac{\ell}{\sqrt{\ell_i^2 + \alpha^2}} \right) + \beta,
  \label{eqn:fit_solid_angle}
\end{equation}
where the fitted parameters are $A=7.27 \pm 0.04$ and $\beta=(6.7 \pm 0.2) \cdot 10^{-4}$ (fig. \ref{fig:distance_attenuated_counts}).
\par
A pair plasma diffusion-like source distribution of $f(x)= \delta(x=0)\delta(y-y_0) + \delta(r-r_0)$, with $y_0=-20$ cm and $r_0=7$ cm, can be simulated with $^{22}\mathrm{Na}$ source $7$ cm off-center on the turntable to simulate the circular emission profile and a stationary $^{22}\mathrm{Na}$ source at $y=-20$ cm to simulate the emission from a limiter.
An equal transport fraction can be simulated by acquiring counts from the same source and for equal time at each source position.  
The emission from an axisymmetric source coaxial with the gamma detector results in an approximately equal count on all detectors with differences up to 13\% due to variance in the detector efficiency (fig. \ref{fig:distance_attenuated_counts} b).
Measurements of a known source located at the center can be used to calibrate these count differences.
The emission from the `limiter' source at $y_0=-20$ cm can be identified by determining the count fraction expected on each detector (red dashed in fig. \ref{fig:distance_attenuated_counts} c). 
The residual difference between the expected counts and the actual counts above the adjusted axisymmetric counts is 3\%.
This demonstrates that the fractional single photon counts on a detector array can identify the counts from a localized source in the presence of an axisymmetric background.
The localized emission rate can be estimated from a least-squares fit to the detector photon counts.
\subsection{Tomographic reconstruction of volumetric coincidence sources\label{sec:Tomo}}
\begin{figure}
  \centering{\includegraphics{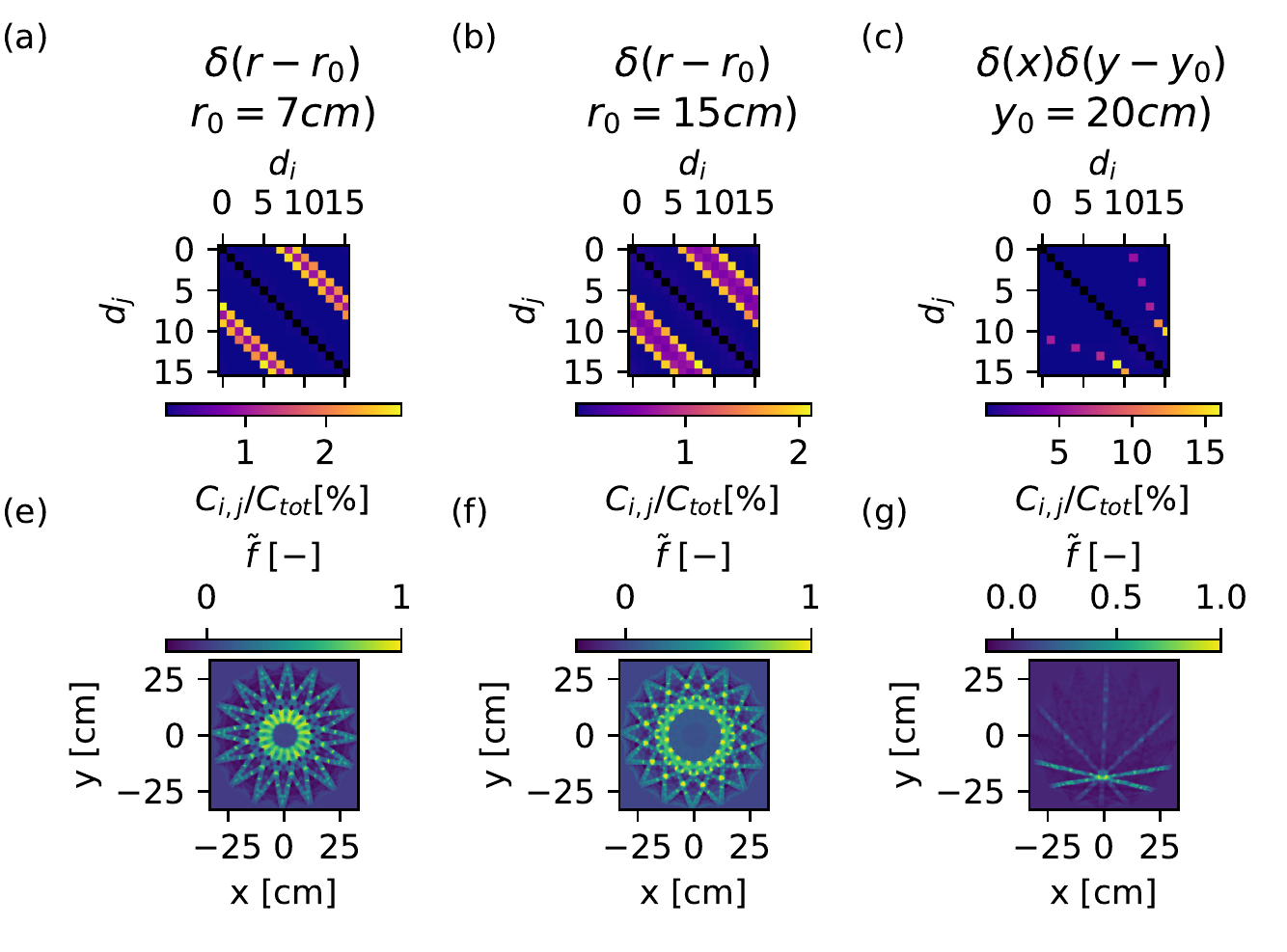}}
  \caption{Coincident counts and 2-$\gamma$ emission profile reconstructions of source distribution functions noted in titles.
(a-c) Relative fraction of coincident counts on each detector pair $i, j$.
(d-f) Reconstruction of emission profile calculated by multiplying coincident count vector with inverse of system response matrix A.}
\label{fig:coin}
\end{figure}
Equation (\ref{eqn:detection_equation}) gives a set of linear equations that can solved for the emission source distribution.
For coincident counts $C_{ij}$ of detectors $i$ and $j$ we can express the equation set as a matrix multiplication with a row for each detector.
Denoting matrices in bold,
\begin{equation}
  \label{eqn:fov_integral}
  \textbf{C} = \textbf{A} \textbf{f}.
\end{equation}
\textbf{A}, the system response function incorporates effects such as the sensitivity of the detectors, non-collinearity due to pair momentum, scattering and attenuation \citep{Baker1992}.
There are several strategies for inverting these equations and choices for basis functions for the reconstructed distribution, e.g. sinusoids in the filtered back projection algorithm \citep{Hobbie2015}.
\par
Fig. (\ref{fig:coin} a-c) shows the counts on the lines of response matrix for distribution functions simulating the diffusion onto the magnet ($f=\delta(r-r_0)$ with $7$ $\mathrm{cm}$), the pair plasma ($f=\delta(r-r_0)$ with $r_0=15\mathrm{cm})$) and the limiter ($f=\delta(x)\delta(y-y_0)$ with $y_0=-20$ $\mathrm{cm})$).
There are no coincident counts on the same detector $i=j$, since the dead time is longer than the coincidence interval (24 ns).
Axisymmetric distributions appear as off-diagonal lines in the count matrices.
The limited number of diagonals with 16 detectors indicates that the radial resolution is limited.
To invert the coincident counts, we estimate the system response matrix by tallying the intersections of uniformly discretized in-plane emission angles ($10^5$ angles) with the detectors.
This is done for point sources at $96$x$96$ discretized locations inside the $20$ cm confinement space.
The resulting matrix is sparse ($>95$\% of entries zero) and can be pseudo-inverted \citep{Penrose1955MPCPS} with an SVD factorization $A=U S V^{\ast} \Rightarrow A^+=V S^{-1} U^{\ast}$, where U is a unitary matrix with m$\times$m, S a diagonal matrix, and V the conjugate transpose of a unitary matrix with n$\times$n elements.
The 100 largest values are used for this pseudo-inversion. 
The source distribution can then be reconstructed with the dot product of the pseudo-inverse $\textbf{A}^+$ and the detector counts $\textbf{C}$, 
\begin{equation}
\textbf{f} = \textbf{A}^+ \cdot \textbf{C}.
\end{equation}
\begin{figure}
  \label{fig:coin_profiles}
  \centering{\includegraphics{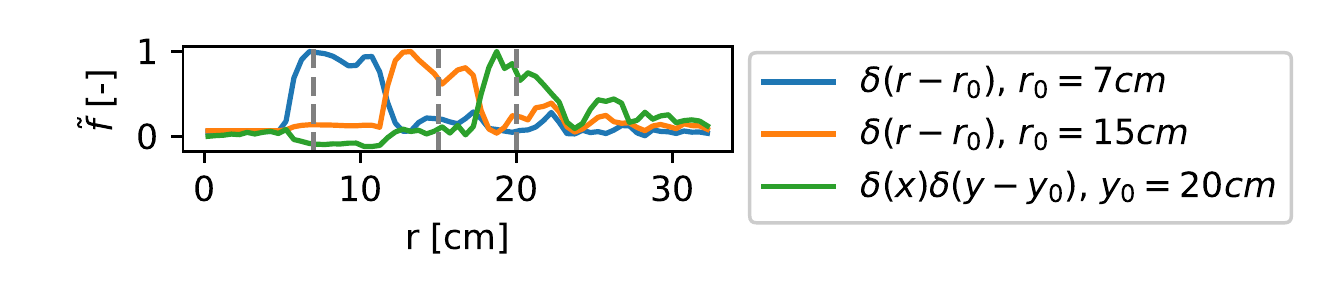}}
  \caption{Radial profiles from tomographic reconstruction of the source distribution function noted in legend.
The blue and orange lines are toroidal averages and the green line is a toroidal sum of the 2D reconstruction of the respective source distribution.}
\end{figure}
Fig. (\ref{fig:coin} d-f) shows the reconstructions based on the count matrices.
Reasonable reconstructions require about $10^6$ coincident counts, which is reasonable for a large part of the density-temperature space (fig. \ref{fig:simulated_emission_fractions} a) and the solid angle coverage (fig. \ref{fig:solid_angle} b).
Artifacts (resembling the coincident solid angle coverage fig. \ref{fig:solid_angle} b) are visible due to the coarse sampling of the area with LORs between only 16 detectors.
Toroidally averaging the 2D reconstructions gives radial profiles removing the artifacts and requiring fewer counts, $\sim 10^4$ (fig. \ref{fig:coin_profiles}).   
The resolution is limited by the number of detectors to a few cm as indicated by the width of the peaks in the radial profiles.
The results shown here demonstrate that the coincident counts from a gamma detector array can be used to reconstruct an emission profile similar to that expected from a magnetically confined pair plasma.
Volumetric emission can be differentiated from the dominant emission due to diffusion with appropriate placement of LORs.
\section{Conclusion}
Magnetically confined pair plasma will exhibit both volumetric and localized annihilation.
We have demonstrated two techniques for diagnosing this emission by imitating matter-antimatter plasma emission with stationary and rotating $\beta^+$ emitters on a turntable. 
Transport processes result in localized annihilation where field lines intersect walls, limiters, or internal magnets.
These localized sources can be identified by the fractional $\gamma$ counts on spatially distributed detectors.
Constraining the annihilation rate for these events may provide insight into the dominant transport processes in magnetically confined pair plasma.
Direct and positronium-mediated annihilation result in overlapping volumetric $\gamma$ sources, and the 2-$\gamma$ emission from these volumetric sources can be tomographically reconstructed from coincident counts.
Compared to the simulation with a $\beta^+$ source the matter-antimatter plasma will present the challenge of disentangling the contributions of Ps formation and direct annihilation.
Estimating the ratio of $2$-$\gamma$ to $3$-$\gamma$ emission with the measured energy spectrum as well as LORs with a long path along the wall may help disentangle the contribution of Ps formation and direct annihilation.
\section*{Acknowledgements}
This work was supported by the European Research Council (ERC-2016-ADG no. 741322), the Deutsche Forschungsgemeinschaft (Hu 978/15, Hu 978/16, Sa 2788/2), the Max Planck Institute for Plasma Physics, the NIFS Collaboration Research Program, Japan Society for the Promotion of Science (JSPS Grants-in-Aid for Scientific Research 25707043 and 16KK0094), the Helmholtz Association Postdoc Programme (E.V.S.).
The work at UCSD is supported by US DOE grant DE-SC0019271 and the UCSD Foundation.
JvdL acknowledges support from the Alexander von Humboldt Foundation.
The authors acknowledge Christoph Hugenschmidt, Francesco Guatieri, Vassily Burwitz, and Max Viehl for lending the $^{22}\mathrm{Na}$ sources and discussions of the CAEN digitizers and FPGA firmwares.
\bibliographystyle{jpp}
\bibliography{bibliography.bib}

\makeatletter
\def\fps@table{h}
\def\fps@figure{h}
\makeatother

\section*{Appendix}

\section*{Annihilation rates}
The rates of annihilation are defined here with equations for $\Gamma$, the annihilation rate of a single positron.
\subsection*{Direct annihilation}
Direct collisions of free positrons and electrons result in annihilation.
In the non-relativistic-limit, the direct annihilation rate $\Gamma_d$ for a positron inside an electron cloud \citep{Crannell1976TAJ} is given by
\begin{equation}
\label{eqn:direct_annihilation_rate}
\Gamma_d = \pi r_0^2 c n_e J(a),
\end{equation}
where $r_0$ is the classical radius of an electron (or positron), $c$ is the speed of light, $n_e$ is the electron density, and $J(r)$ is a Coulomb-attraction enhancement factor defined as
$J(a)=\frac{4a}{\pi^{1/2}} \int_0^\infty \frac{x e^{-x}}{1-e^{-a/x}} \diff x $,
where $a=\sqrt{2\pi^2 Ry/kT}$, $k$ is Boltzmann's constant, $T$ is the pair temperature, and $Ry$ is a Rydberg unit of energy ($Ry=hcR_\infty$).
$J(a) \sim 1$ for temperatures above $100$ eV so that the annihilation rate only depends weakly on temperature.
For temperatures below $100$ eV $J(a)$ and $\Gamma_d$ scale as $\sqrt{T}$.
\subsection*{Radiative recombination}
Annihilation also occurs through decay of short-lived bound states of an electron and positron, positronium (Ps).
Ps forms through interactions with a third particle.
The third particle is a photon in the radiative recombination process, another electron or positron in the three-body recombination process, or a bound electron in a charge-exchange process.
The rate of radiative recombination of a positron in an electron gas, $\Gamma_r$, can be expressed in terms of a modified hydrogenic radiative recombination coefficient \citep{Gould1989TAJ}, $\alpha_{H}$, where the mass of the electron $m_e$ is replaced by $m_e/2$,
\begin{equation}
\label{eqn:radiative_recombination_rate}
  \Gamma_r = n_e  \alpha_{Ps} = n_e \alpha_{H}(m_e \rightarrow m_e/2).
\end{equation}
The Ps radiative coefficient is then
\begin{equation}
  \alpha_{Ps} = 256 \cdot 3^{-3/2} \alpha^3 \pi \left( \frac{\hbar^2}{m_e e^2} \right)^2 \sqrt{\frac{4kT}{\pi m_e}} \frac{Ry}{2 kT} \phi \left(\frac{Ry}{2kT} \right) \bar{g} \left(\frac{Ry}{2kT} \right),
\end{equation}
where $\alpha$ is the fine-structure constant, $\hbar$ is the reduced Planck constant, $m_e$ is the electron mass, $Ry$ is a Rydberg unit of energy ($Ry=hcR_\infty$), $\phi(x)$ is a transcendental function that captures the contributions due to formation at principal quantum numbers, and $\bar{g}(x)$ is an averaged Gaunt factor.
$\phi(x)$ as well as values of $g(x)$ are given in Refs. \citep{Gould1989TAJ}, \citep{Gould1972AP}.
The rate of radiative recombination scales as $T^{-0.85}$ for temperatures below $\tilde 50$ eV and exceeds direct annihilation below $59$ eV.
\subsection*{Diffusion due to collisions with neutrals}
For neutral collisions the step size should correspond to the Larmor radius.
The confinement time for positrons in a permanent magnet trap with mean radial distance to the wall of $2.5$ cm has been related to an estimated collision number of 200 \citep{HornStanja2018PRL}.
Scaling the collisions to the $7$ cm mean radial distance of the levitated dipole configuration described in \citep{Stoneking2020JPP} provides an estimate of, $N_{coll}\sim 1800$ collisions needed to traverse the trap.
The diffusion rate due to neutral collisions is then
\begin{equation}
\label{eqn:neutral_transport_rate}
  \Gamma_{N} = \nu/N_{coll},
\end{equation}
where $\nu$ is the collision rate.
The collision rate can be obtained from measurements of the total cross section of positrons with atoms and molecules \citep{Zhou1997PRA, Zecca2006JPB}, e.g. the diffusion due to molecular hydrogen.
\subsection*{Diffusion due to Coulomb collisions in a strongly magnetized pair plasma}
\par
Coulomb collisions between charged particles result in diffusion of positrons and subsequent annihilation on material surfaces with a rate given by,
\begin{equation}
\label{eqn:coulomb_transport_rate}
  \Gamma_{cc}=\frac{D_{cc}}{x^2},
\end{equation}
where $x$ is the mean length to the limiter and $D_{cc}$ the Coulomb collision diffusion coefficient.
Theory \citep{Dubin1998POP, Dubin1997PRL} and observations \citep{Anderegg1997PRL} in non-neutral plasma suggest the diffusion coefficient is enhanced for collisions with an impact parameter larger than the Larmor radius, $\rho > r_L$, while the effect of collisions with $\rho<r_L$ can be described by classical diffusion,
\begin{equation}
  \label{eqn:diffusion_split}
D_{cc} = D_{clas} + D_{mag}.
\end{equation}
The strongly magnetized diffusion coefficient, $D_{mag}$, applies to impact factors larger than the Larmor radius, $r_L < \rho < \lambda_D$, where magnetic moment conservation makes collisions appear as $\vec{E} \times \vec{B}$ drifts due to the electric field of the particles,
\begin{equation}
\label{eqn:strongly_magnetized_diffusion}
D_{mag}=2 \sqrt{\pi} ln(\lambda_D/r_L)ln(u/(\nu_cu^2\sqrt{\lambda_D r_L})^{1/3}) \nu_c r_L^2,
\end{equation}
where $\nu_c$ is the classical plasma collision frequency, and $u$ the thermal velocity.
The diffusion can be further enhanced by correlated collisions \citep{Dubin1998POP}.
For magnetized diffusion calculations, the magnetic field will be set to $B=1$ $\mathrm{T}$.
At low temperatures ($< 30$ K) the correlation between particles can become extreme, effectively forming magnetobound states between electrons and positrons leading to a large increase in transport \citep{Aguirre2017POP}.
The classical diffusion has its usual form,
\begin{equation}
  \label{eqn:classical_diffusion}
D_{clas}=\frac{4}{3}\sqrt{\pi} \nu_c r_L^2. 
\end{equation}
\subsection*{Positronium - Excited States and Lifetimes}
Radiative recombination can form oPs and pPs at excited states.
The production rates $\Gamma_n$ for each principal quantum number $n$ are given for hydrogen in \citep{Gould1972AP} and can be adjusted for Ps with equation (\ref{eqn:radiative_recombination_rate}).    
Most of the states will de-excite to one of the S states (angular momentum $l=0$) before annihilating \citep{Gould1989TAJ}.
The lifetimes of the excited Ps states are assumed to be twice those of atomic decay in hydrogen \citep{Gould1972AP}.
The cascades of Ps states used in the calculations in this paper are given with their formation fraction and total lifetime in tables \ref{tab:pPs_appendix} and \ref{tab:oPs_appendix}.
\begin{table}
\begin{center}
{\tabcolsep1pc\begin{tabular}{@{}cccc@{}}
state cascade & principal number & lifetime\\
 & production rate fraction & \\
$4S$ & $\frac{1}{16} \Gamma_4$ & $8$ $\mathrm{ns}$\\
$4P \rightarrow 3S$ & $\frac{1}{32} \Gamma_4$ & $0.7$ $\mathrm{\mu s} + 3.4$ $\mathrm{ns}$\\
$3S$ & $\frac{1}{9} \Gamma_3$ & $3.4$ $\mathrm{n s}$\\
$4P \rightarrow 2S$ & $\frac{1}{32} \Gamma_4$ & $0.2$ $\mathrm{\mu s}+1$ $\mathrm{ns}$\\
$4D \rightarrow 3P \rightarrow 2S$ & $\frac{5}{64} \Gamma_4$ & $0.3$ $\mathrm{\mu s}+91$ $\mathrm{ns} + 1$ $\mathrm{ns}$\\
$3P \rightarrow 2S$ & $\frac{1}{6} \Gamma_3$ & $91$ $\mathrm{ns}+1$ $\mathrm{ns}$\\
$2S$ & $\frac{1}{4} \Gamma_2$ & $1$ $\mathrm{ns}$\\
$4P \rightarrow 1S$ & $\frac{1}{32} \Gamma_4$ & $29$ $\mathrm{ns} + 125$ $\mathrm{ps}$\\
$4D \rightarrow 3P \rightarrow 1S$ & $\frac{5}{64} \Gamma_4$ & $0.3$ $\mathrm{\mu s}+12$ $\mathrm{ns} + 125$ $\mathrm{ps}$\\
$4D \rightarrow 2P \rightarrow 1S$ & $\frac{5}{32} \Gamma_4$ & $98$ $\mathrm{ns}+3$ $\mathrm{ns} + 125$ $\mathrm{ps}$\\
$4P \rightarrow 3D \rightarrow 2P \rightarrow 1S$ & $\frac{3}{32} \Gamma_4$ & $6.7$ $\mathrm{\mu s}+31$ $\mathrm{ns}+3$ $\mathrm{ns} + 125$ $\mathrm{ps}$\\
$4F \rightarrow 3D \rightarrow 2P \rightarrow 1S$ & $\frac{7}{16} \Gamma_4$ & $0.15$ $\mathrm{\mu s}+31$ $\mathrm{ns} + 3$ $\mathrm{ns} + 125$ $\mathrm{ps}$\\
$3P \rightarrow 1S$ & $\frac{1}{6} \Gamma_3$ & $12$ $\mathrm{ns} + 125$ $\mathrm{ps}$\\
$3S \rightarrow 2P \rightarrow 1S$ & $\frac{5}{9} \Gamma_3$ & $31$ $\mathrm{ns} + 3$ $\mathrm{ns} + 125$ $\mathrm{ps}$\\
$2P \rightarrow 1S$ & $\frac{3}{4} \Gamma_2$ & $3$ $\mathrm{ns} + 125$ $\mathrm{ps}$\\
$1S$ & $\Gamma_1$ & $125$ $\mathrm{p s}$\\
\end{tabular}}
\end{center}
\caption{Population of S states and their respectives lifetimes after pPs formation in states up to n = 4.}
\label{tab:pPs_appendix}
\end{table}

\begin{table}
\begin{center}
{\tabcolsep1pc\begin{tabular}{@{}cccc@{}}
state cascade & principal number & lifetime\\
 & production rate fraction & \\
$4S \rightarrow 3P \rightarrow 2S$ & $\frac{1}{64} \Gamma_4$ & $1.1$ $\mathrm{\mu s}+91$ $\mathrm{ns} + 1.1$ $\mathrm{\mu s}$ \\
$4S \rightarrow 2S$ & $\frac{1}{32} \Gamma_4$ & $0.2$ $\mathrm{\mu s}+1.1$ $\mathrm{\mu s}$\\
$4D \rightarrow 3P \rightarrow 2S$ & $\frac{5}{64} \Gamma_4$ & $0.3$ $\mathrm{\mu s}+91$ $\mathrm{ns} + 1.1$ $\mathrm{\mu s}$\\
$3P \rightarrow 2S$ & $\frac{1}{6} \Gamma_3$ & $91$ $\mathrm{ns} + 1.1$ $\mathrm{\mu s}$\\
$2S$ & $\frac{1}{4} \Gamma_2$ & $1.1$ $\mathrm{\mu s}$\\
$4S \rightarrow 3P \rightarrow 1S$ & $\frac{1}{64} \Gamma_4$ & $1.1$ $\mathrm{\mu s}+12$ $\mathrm{ns} + 142$ $\mathrm{ns}$\\
$4S \rightarrow 2P \rightarrow 1S$ & $\frac{1}{32} \Gamma_4$ & $0.8$ $\mathrm{\mu s}+3$ $\mathrm{ns} + 142$ $\mathrm{ns}$\\
$4P \rightarrow 3S \rightarrow 2P \rightarrow 1S$ & $\frac{1}{32} \Gamma_4$ & $0.7$ $\mathrm{\mu s}+0.3$ $\mathrm{\mu s}+3$ $\mathrm{ns} + 142$ $\mathrm{ns}$\\
$4P \rightarrow 3D \rightarrow 2P \rightarrow 1S$ & $\frac{3}{32} \Gamma_4$ & $6.7$ $\mathrm{\mu s}+31$ $\mathrm{ns}+3$ $\mathrm{ns} + 142$ $\mathrm{ns}$\\
$4P \rightarrow 1S$ & $\frac{1}{32} \Gamma_4$ & $29$ $\mathrm{ns} + 142$ $\mathrm{ns}$\\
$4D \rightarrow 3P \rightarrow 1S$ & $\frac{5}{64} \Gamma_4$ & $0.3$ $\mathrm{\mu s}+12$ $\mathrm{ns} + 142$ $\mathrm{ns}$\\
$4D \rightarrow 2P \rightarrow 1S$ & $\frac{5}{32} \Gamma_4$ & $98$ $\mathrm{ns}+3$ $\mathrm{ns} + 142$ $\mathrm{ns}$\\
$4F \rightarrow 3D \rightarrow 2P \rightarrow 1S$ & $\frac{7}{16} \Gamma_4$ & $0.15$ $\mathrm{\mu s}+31$ $\mathrm{ns}+3$ $\mathrm{ns} + 142$ $\mathrm{ns}$\\
$3S \rightarrow 2P \rightarrow 1S$ & $\frac{1}{9} \Gamma_3$ & $0.3$ $\mathrm{\mu s}+3$ $\mathrm{ns} + 142$ $\mathrm{ns}$\\
$3P \rightarrow 1S$ & $\frac{1}{6} \Gamma_3$ & $12$ $\mathrm{ns} + 142$ $\mathrm{ns}$\\
$3D \rightarrow 2P \rightarrow 1S$ & $\frac{5}{9} \Gamma_3$ & $31$ $\mathrm{ns} + 3$ $\mathrm{ns} + 142$ $\mathrm{ns}$\\
$2P \rightarrow 1S$ & $\frac{3}{4} \Gamma_2$ & $3$ $\mathrm{ns} + 142$ $\mathrm{ns}$\\
$1S$ & $\Gamma_1$ & $142$ $\mathrm{ns}$\\
\end{tabular}}
\end{center}
\caption{Population of S states and their respective lifetimes after oPs formation with states up to n = 4.}
\label{tab:oPs_appendix}
\end{table}

\end{document}